\documentclass[preprint, double-spaced,showkeys,showpacs]{revtex4}
\usepackage{graphicx}
\usepackage{amsmath,bbm}
\usepackage{amssymb,bm}
\usepackage{array}
\usepackage[linktocpage=true,colorlinks=true,linkcolor=blue,citecolor=blue]{hyperref}
\begin{document}
\title{Heavy Quarkonia in a Potential Model: Binding Energy, Decay Width, and Survival Probability}
\author{P. K. Srivastava$^1$\footnote{prasu111@gmail.com}}
\author{O. S. K. Chaturvedi$^2$}
\author{Lata Thakur$^3$\footnote{latathakur@prl.res.in}}
\affiliation{$^1$Department of Physics, Indian Institute of Technology 
Ropar, Rupnagar-140001, INDIA}
\affiliation{$^2$ Department of Physics, Institute of Science, Banaras Hindu University, Varanasi-221005, INDIA}
\affiliation{$^3$ Theory Division, Physical Research Laboratory,
	Navrangpura, Ahmedabad 380 009, INDIA}
\begin{abstract}
Recently a lot of progress has been made in deriving the heavy quark potential within a QCD medium.
In this article we have considered heavy quarkonium in a hot quark gluon plasma phase. 
The heavy-quark potential has been modeled properly for short as well as long distances. 
The potential at long distances is modeled as a QCD string which is screened 
at the same scale as the Coloumb field. We have numerically solved the  
1+1-dimensional Schrodinger equation for this potential and obtained the eigen wavefunction 
and binding energy for the $1S$ and $2S$ states of charmonium and bottomonium. Further, 
we have calculated the decay width and dissociation temperature of quarkonium states in the QCD plasma. Finally, we have used our recently proposed 
unified model with these new values of decay widths to calculate the survival probability 
of the various quarkonium states with respect to centrality at relativistic heavy ion collider (RHIC) and large hadron collider (LHC) energies. This study provides a unified, consistent and comprehensive description of spectroscopic properties of various quarkonium states at finite temperatures along with their nuclear modification factor at different collision energies.
\\

\end{abstract}

\maketitle 
\section{Introduction}
\noindent
Heavy quarkonium production and suppression was one of the earliest proposed tool to study the properties of the medium created in heavy ion collisions. In mid 80's, Matsui and Satz~\cite{Matsui:1986dk}has proposed theoretically that quarkoinum suppression is the signal of the possible creation of quark gluon plasma (QGP) in collision experiments. From there onward, the physical picture of quarkonium dissociation in a thermal medium has undergone various theoretical and experimental refinements~\cite{Brambilla:2010cs}. Recent experimental observations suggest that the charmonium suppression in QCD plasma is not the result of a single mechanism, but is a complex interplay of various physical processes. Heavy quarkonia ($ Q\bar{Q} $) has a special edge over many other proposed tools since the heavy mass scale ($m_{J/\psi}=3.1$ GeV  for $J/\psi$ and $m_\Upsilon=9.2$ GeV for $\Upsilon$) makes this system possible for analytical treatment theoretically. On the other side, decay of heavy quarkonia via dileptonic channel lead to relatively clean signal which can be precisely measured experimentally.  

We can get the physical insight of the medium dependence by analyzing the behavior of spectral function of heavy quarkonium. The two useful approaches to study the production and suppression via spectral function of heavy quarkonium are potential method and lattice approach~\cite{Karsch:2000gi,Mocsy:2004bv,Wong:2004zr,Cabrera:2006wh,Mocsy:2007jz,Alberico:2007rg,Mocsy:2007yj,deForcrand:2000akx}. As we all know that lattice QCD method is first principle tool to study the properties and behavior of heavy quarkonium thus none of the potential method can be alternative to this approach. However, the lattice observations are suffering from discretization effects and statistical errors. In this scenario, potential models can be utilized to surve the purpose. As we now know that the problem of heavy quark bound state at zero temperature involves different energy scales, i.e., hard scale, which is the mass $m_Q$ of heavy quark, soft scale which is inverse size $m_Qv\sim 1/r$ of bound state and ultrasoft scale, which is the binding energy $mv^{2}\sim \alpha_{s}/r$. After integrating out the hard scale modes, one obtains an effective field theory non-relativistic QCD (NRQCD)~\cite{Bodwin:1994jh,Bodwin:2010py}. Subsequently, integrating out the modes related with the inverse size scale, potential NRQCD (pNRQCD) appears~\cite{Brambilla:1999xf,Brambilla:2004jw}. In this pNRQCD, the heavy quark-antiquark pair in singlet and octet state are included via dynamical singlet and octet fields (or potentials). 

The generalization of this approach at finite temperature involves three different thermal scales : $T$, $gT$ and $g^{2}T$. In the static limit and if the binding energy is larger than the temperature $T$, the derivation of pNRQCD proceed in the same way as in zero temperature theory and heavy quark potential is not affected by the medium. However, the bound state properties can be affected through interaction of bound states with the ultrasoft gluons of the medium. The main effect of this interaction is the reduction of binding energy of the heavy quark bound state and emergence of a finite thermal width. In second case when one of the thermal scales is higher than binding energy, the singlet and octet potential become temperature dependent and will acquire an imaginary part~\cite{Brambilla:2008cx}. It is important to state here that the real part of the potential leads to colour screening while imaginary part of the potential introduces the Landau damping to the heavy quark bound states~\cite{Laine:2006ns,Thakur:2013nia,Margotta:2011ta,Strickland:2011aa}.

Another observation from numerical lattice calculations 
show the crossover type of deconfinement transition from hadron gas to QGP~\cite{Karsch:2006xs}. Thus, we can expect some non-perturbative effects such as non-vanishing string tension in heavy quark-antiqurk potential above the critical temperature $T_{C}$ as well.
So it is reasonable to assume the string term above $T_{C}$~\cite{Cheng:2008bs,Maezawa:2007fc,Andreev:2006eh}.
Further this potential should also incorporate the effect of Landau damping induced thermal width by calculating the imaginary part of the potential.
In the recent years, the real~\cite{Agotiya:2008ie,Thakur:2012eb} and imaginary parts~\cite{Thakur:2013nia,Thakur:2016cki} of the heavy quark potential have been
calculated by modifying both the perturbative and non-perturbative terms of the Cornell potential in the static as well as in a moving medium.
The complex static interquark
potential at finite temperature has also been derived in Ref.~\cite{Burnier:2015nsa} by considering both the Coulombic and linear string terms.
One can calculate the dissociation coefficient at a given temperature by solving the Schr\"odinger equation using the modified heavy quark complex potential~\cite{Agotiya:2008ie,Thakur:2012eb,Thakur:2013nia,Agotiya:2016bqr,Ganesh:2014lha}. 
Recently we have constructed a unified model for charmonium suppression in Ref.~\cite{Singh:2015eta}. Here we want to incorporate this modified heavy quark potential from Ref.~\cite{Agotiya:2008ie,Thakur:2013nia} in unified model to calculate the survival probability of heavy quarkonium 
states. The survival probability of heavy quarkonium states has been studied recently in Refs.~\cite{Patra:2009qy,Kakade:2015xua}. Specially we will focus here on the double ratio of two states of charmonium since most of the suppression models are failed to reproduce the suppression pattern of this double ratio. 

In this article we have modified our unified model to properly include perturbative as well as nonperturbative effects on quarkonium suppression. We have constructed this model based on the kinetic approach whose original ingredients
was given by Thews et al.~\cite{Thews:2005fs}. In this approach, there are two terms written 
on the basis of Boltzmann kinetic equation as shown in subsection \ref{D}. First term, which we call as dissociation term, 
includes the dissociation process like gluo-dissociation and collisional damping. The second 
term (formation term) provides the (re)generation of $J/\psi$ due to the recombination of 
charm-anticharm quark. These two terms compete over the entire temporal evolution of the QGP 
and we get the multiplicity of finally survived quarkonia at freezeout temperature. To define the dynamics of 
the system created in the heavy ion collisions, we have used the 1+1 dimensional viscous hydrodynamics.
Here we have included only the shear viscosity and neglected the bulk viscosity. We have also suitably 
incorporated the overall feed-down correction from the higher states to the low-lying states.  Rest of the paper is organised as follows : In section \ref{secII}, model formulation, we have provided four subsections which discuss briefly about modified heavy quark potential at finite temperature, binding energy, decay width and calculation of survival probability, respectively. Further in section \ref{secIII}, we have presented our results along with their discussions. In the end of this section, we have also summarized our present work.
 
\section{Model Formalism}
\label{secII}
\subsection{Heavy Quark Complex Potential}
\label{A}
In this section we discuss about the heavy quark-antiquark potential which have both the columbic and string-like parts. Authors in Ref.~\cite{Laine:2006ns} have derived the static potential between heavy quark-antiquark pair at finite temperature by defining a suitable gauge-invariant Green's function and computing it to first order in hard thermal loop (HTL) resummed perturbtion theory. In medium both the columbic and string-like part of potential receive modification. Further complications arise from the fact an imaginary part of the potential arises due to the presence of scattering of light medium degrees of freedom with the color string spanning in between the heavy quarks and antiquarks. It has been pointed out that the physics of the finite width originates from the Landau damping of low-frequency gauge fields. Further it has been studied non-perturbatively by making use of the classical approximation. In the view of above observations, a meaningful description of the relavant physics of quarkonium must therefore consist both the effects of screening of the real part of potential and imaginary part of the potential. There are several efforts to derive and/or phenomenologically construct HQ potential which can be used as an input in the quarkonium suppression models~\cite{Laine:2006ns,Burnier:2015nsa,Krouppa:2018,Blaizot:2016,Rothkopf:2012,Burnier:2017}. The standard Polyakov loop correlator is fail to reproduce the expected Debye-screening potential at asymptotically large distances. Many modified descriptions of Polyakov loop correlator are affected by gauge ambiguities. There are studies based on generalized Gauss law and further its combination with the characterization of in-medium effects through the perturbative HTL permittivity. The use of Gauss law, non-local concept leads to a self-consistent descriptions of both screening and damping effects. Recently, direct lattice determination of the quarkonium spectral function have been attempted~\cite{deForcrand:2000akx}. However, these calculations are again plagued by the model assumptions as there are very finite number of points in time direction and data is of statisctical nature in these lattice studies. On the other side, similar observables have been calculated for strongly-coupled $N=4$ super Yang-Mills theory through AdS/CFT correspondence~\cite{Ali:2014,Patra:2015qoa,Liu:2007,Fadafan:2013,Tahery:2017}. In these derivations, static potentials in real time can be calculated by computing the standard Wilson loop in Euclidean spacetime, and then carry out the analytic continuation.  In other words, the expectation value of a particular timelike Wilson loop defines the potential between a static quark and antiquark at finite temperature. 
The medium modified heavy quark potential can be obtained by correcting both the Coulombic (short-distance)
and string (long-distance) terms, not its Coulomb term alone, with a dielectric function encoding the effects of the
deconfined medium as discussed in Refs.~\cite{Agotiya:2008ie,Thakur:2012eb,Thakur:2013nia}.
In the literature only
a screened Coulomb potential was assumed above critical temperature ($ T_c $) and the non-perturbative (string) term 
was usually overlooked (assumed zero), was certainly
worth investigation.
Recent lattice results indicate the phase transition in
full QCD appears to be a crossover rather than a 
phase transition with the related singularities in thermodynamic observables as discussed in introduction section. The effects of string tension between
the $ Q\bar{Q} $ pairs should not be ignored beyond $ T_c $. Therefore, it is 
important to incorporate the string term while
setting up the criterion for the dissociation. In our approach, we make the assumption that medium potential can be derived from 
the vacuum potential by multiplying it with a field-theoretically determined complex permittivity in momentum space. It is possible 
to reproduce the real and imaginary part of the corresponding in-medium potential in by using the  hard thermal loop permittivity.
The real part of the medium modified heavy quark potential can be written as~\cite{Agotiya:2008ie,Thakur:2012eb}
\begin{equation}
 ReV (r,~T) = -\alpha m_{D}\left(\frac{e^{-\hat r}}{\hat r}+1\right)+\frac{2\sigma}{m_{D}}\left(\frac{e^{-\hat r}-1}{\hat r}+1 \right),
 \label{Repot}
\end{equation}
where $\hat{r}= rm_{D}$ and $\alpha=4/3~\alpha_{s}$ with $\alpha_s $ 
as the one loop running coupling constant given as
\begin{eqnarray}
\alpha_{s}(T)=\frac{g_s^{2}(T)}{4 \pi}=\frac{6 \pi}{\left(33-2 N_{f}\right)\ln \left(\frac{2\pi T}{\Lambda_{\overline{\rm MS}}}\right)}.
\end{eqnarray}
  Here we take $\Lambda_{\overline{\rm MS}}=0.1  $ GeV and the string tension, $ \sigma=0.184 $ GeV$^2$. 
$m_{D}$ is the Debye screening mass which is defined as
\begin{equation}
 m_{D}^{2} = \frac{g^{2}T^{2}}{6}\left(N_{f}+2N_{c}\right),
\end{equation}
with $N_{f}$ and $N_{c}$ as the number of flavours and colours, respectively.\\ 
 In the small $r$ limit, the real part of potential
reduces to the Cornell potential.
\begin{equation}
ReV (r)\approx -\frac{\alpha}{r}+ \sigma r.
\end{equation}
On the other hand, in the large distance limit (where the screening occurs), potential is reduced to a
long-range Coulomb potential with a dynamically
screened-color charge. However, if we compare our $ Q\bar{Q} $ potential (Eq.~\ref{Repot}) with the classical concept of Debye-H\"{u}ckel theory by Digal et.al.~\cite{Digal:2005ht}, we found that in the asymptotic limit $ (r \rightarrow \infty) $, Eq.~(\ref{Repot}) reduces to 
\begin{equation}
ReV (r\rightarrow \infty,~T)= F(\infty,T)=\frac{2\sigma}{m_{{D}}}-\alpha m_{{D}}, 
\end{equation}
whereas in Ref.~\cite{Digal:2005ht} free energy reduces to
 \begin{equation}	F^{\rm{Digal}} (\infty,T)=
\frac{\Gamma(1/4)}{2^{3/2}\Gamma(3/4)}\frac{\sigma}{m_{{D}}}-\alpha m_{{D}},
\end{equation}
here the difference can be seen only in the string term only and may be due to
the treatment of the problem classically or quantum mechanically. Also in the framework of Debye-H\"{u}ckel theory,
Digal et al. employed different screening functions,
$f_c$ and $f_s$ for the Coulomb and string terms, respectively, to obtain
the free energy. Here we have used the same screening scale, $ m_D $ for both the Coulombic and linear terms.
\begin{figure}
\includegraphics[scale = 0.7]{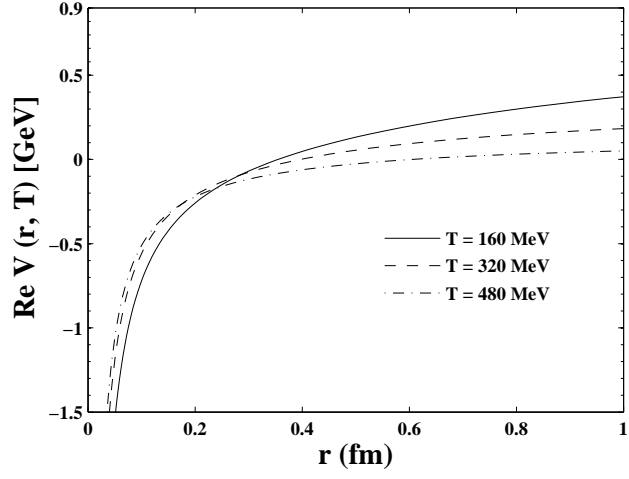}
\caption{Variation of the real part of potential with the separation distance $r$ between the $ Q\bar{Q} $ pair at three different values temperatures, i.e., $T=160,~320$ and $480$ MeV. }
\label{ReV}
\end{figure}
 
The imaginary part of the medium modified heavy quark potential can be calculated in the similar way as in Ref.~\cite{Thakur:2013nia} and is given by
\begin{equation}
ImV(r,~T) = -\alpha T~\phi(m_D r) -\frac{2\sigma T}{m_{D}^{2}}\psi(m_D r),
\end{equation}
where the functions $ \phi(\hat r) $ and $ \psi(\hat r) $ are defined as
\begin{equation}
\phi(\hat r)=2\int_{0}^{\infty}\frac{z~dz}{(z^{2}+1)^{2}}\left(1-\frac{sin~z\hat{r}}{z\hat{r}}\right)
\end{equation}
and 
\begin{equation}
\psi (\hat r) = 2\int_{0}^{\infty}\frac{dz}{z(z^{2}+1)^{2}}\left(1-\frac{sin~z\hat{r}}{z\hat{r}}\right),
\end{equation}
\begin{figure}
	\includegraphics[scale=0.7]{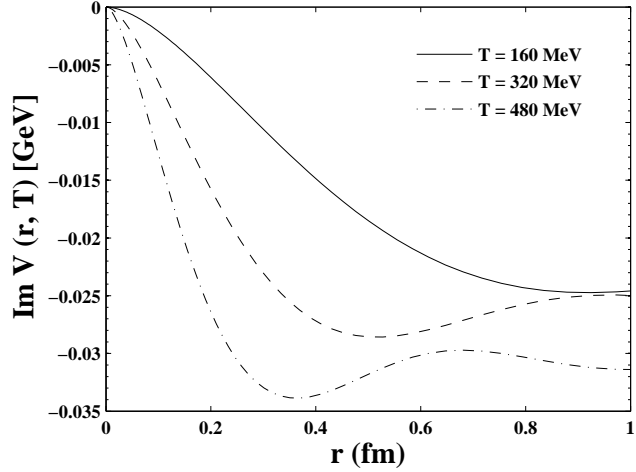}
	\caption{Variation of the imaginary part of potential with the separation distance $r$ at three different values of temperature.}
	\label{ImV}
\end{figure}
In the small $r$ limit, we can expand the potential and at leading logarithmic order in $\hat{r}$ we get
\begin{eqnarray}
Im V (\hat r,T)\approx-\alpha T\frac{ {\hat r^2}}{3}\log\big(\frac{1}{\hat r}\big)
-\frac{2\sigma T}{m_D^2}\left(\frac{\hat r^2}{6}  -\frac{\hat r^4}{60}\right)\log\big(\frac{1}{\hat r}\big).
\label{Impot}
\end{eqnarray}

\begin{figure}
\includegraphics[scale=0.6]{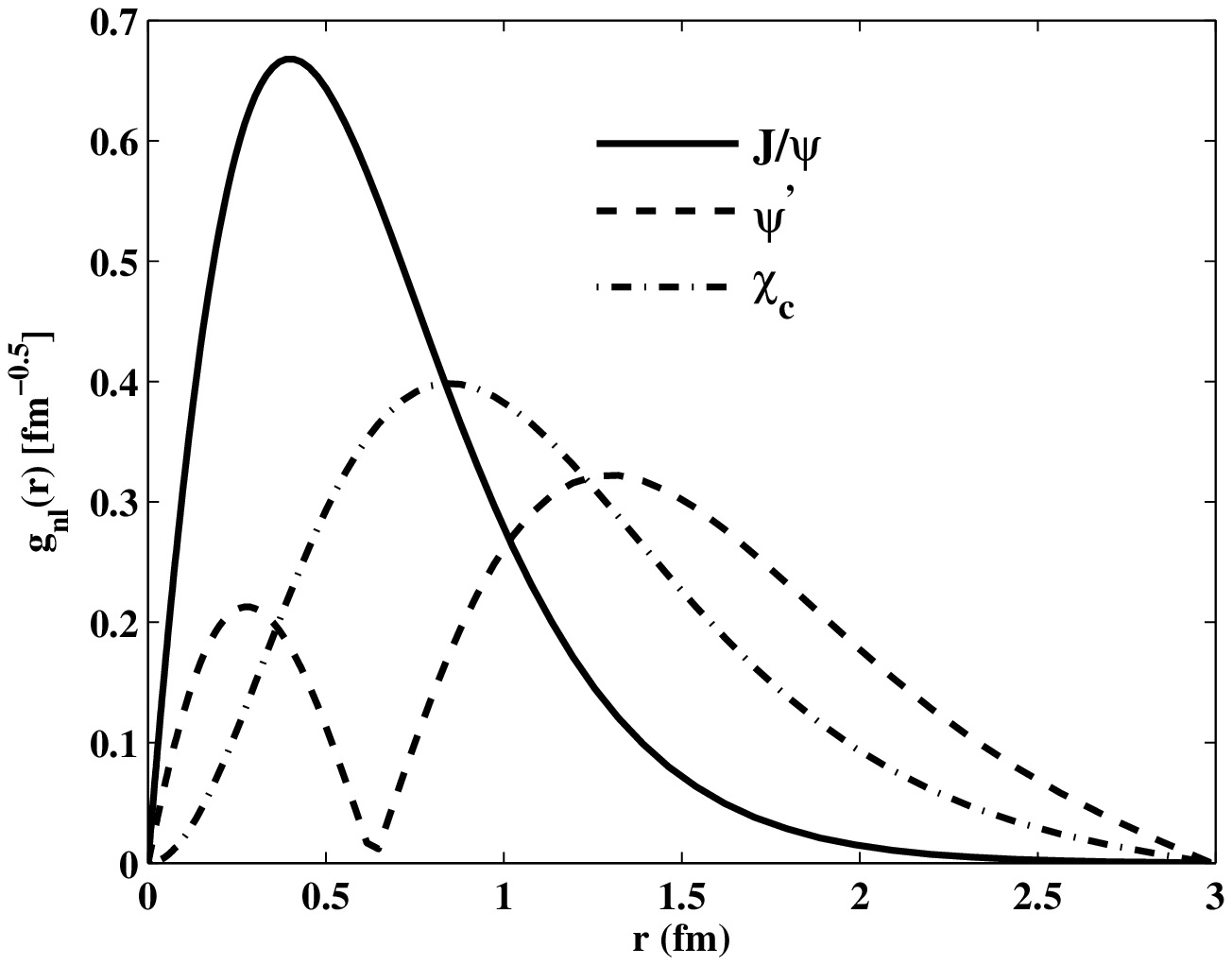}
\caption{The variation of radial part of wavefunction of different charmonia states with respect to $r$. 
Solid curve represents the $1S(J/\psi)$, dashed curve is for $2S(\psi^{'})$ and dash-dotted curve represents
the radial wavefunction of $1P(\chi_{c})$ state.}
\label{fig3}
\end{figure}
\begin{figure}
\includegraphics[scale=0.6]{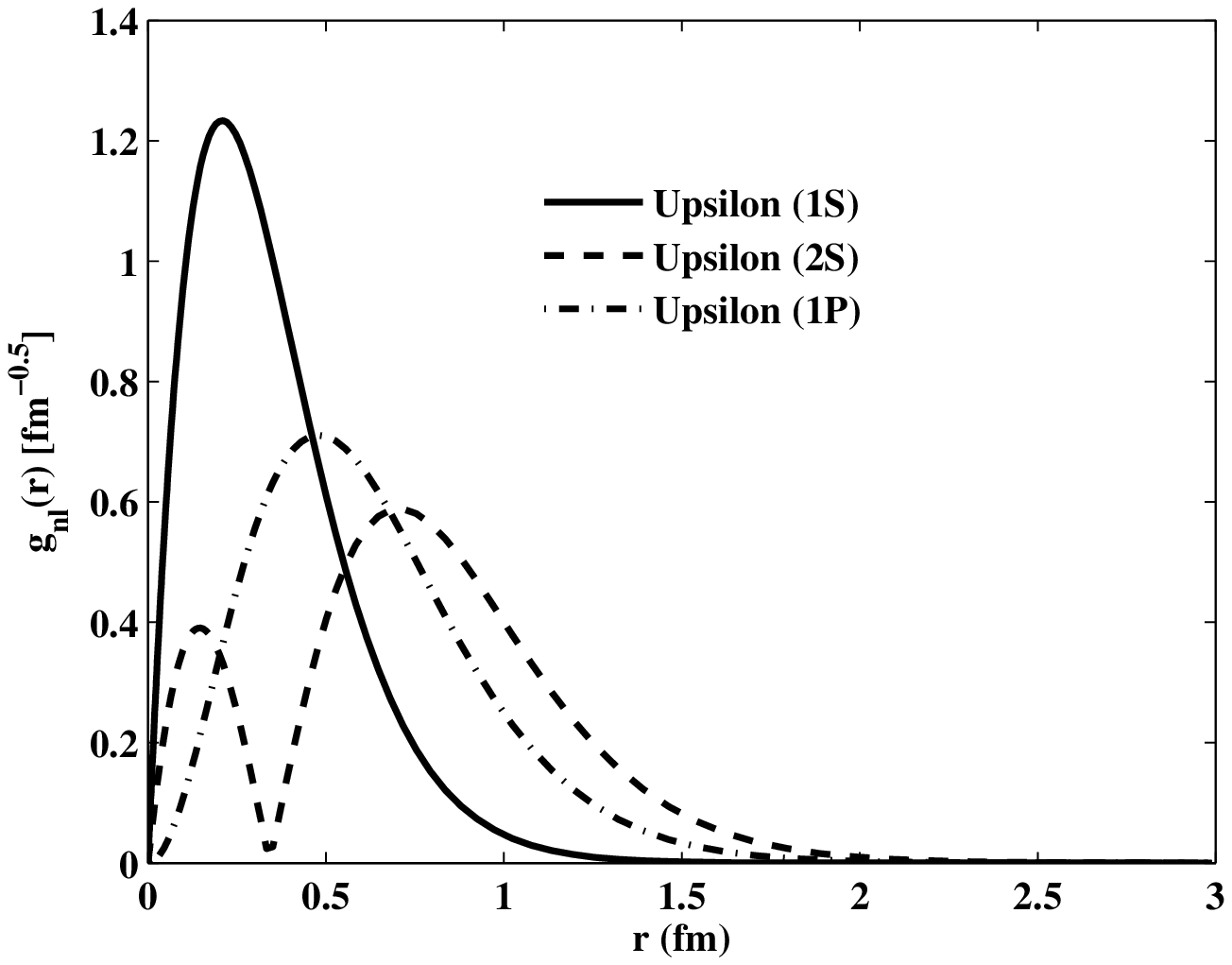}
\caption{The variation of radial part of wavefunction of different bottomonia states with respect to $r$. 
Solid curve represents the $1S(\Upsilon)$, dashed curve is for $2S(\Upsilon)$ and dash-dotted curve represents
the radial wavefunction of $1P(\chi_{b})$ state.}
\label{fig4}
\end{figure}
\begin{figure}
\includegraphics[scale=0.6]{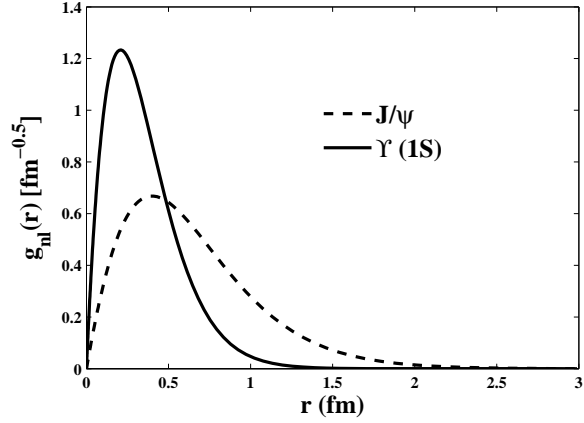}
\caption{Comparison of radial part of wavefunction of $\Upsilon(1S)$ and $J/\psi$ at 
critical temperature $T_{C}=160$ MeV.}
\label{fig5}
\end{figure}
\begin{figure}
\includegraphics[scale=0.6]{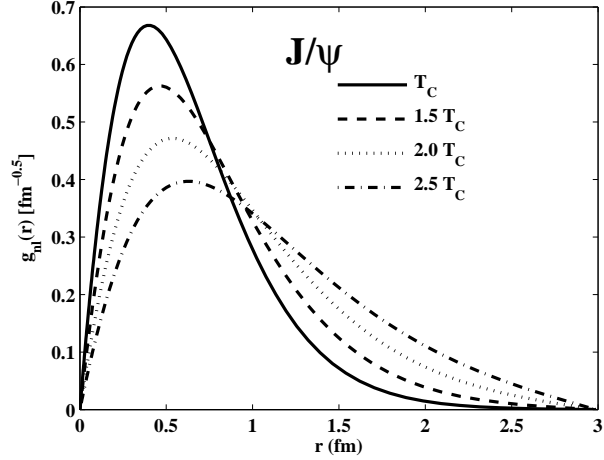}
\caption{Variation of radial wavefunction of $J/\psi$ with respect to $r$ 
at $T=T_{C},~1.5~T_{C},~2.0~T_{C}$ and $2.5~T_{C}$.}
\label{fig6}
\end{figure}

\subsection{Binding Energy}
\label{B}
Binding energy can be calculated by knowing the energy eigen value of different quarkonia state. The binding energy here
is a function of temperature instead of a constant factor. To know the energy eigen value and energy eigen function of different quarkonia states, we have solved the Schrodinger equation
with the real part of heavy quark potential as described above and a angular momentum dependent part. The motivation behind using only the real part 
of the potential in solving Schrodinger equation is the large magnitude of real part over imaginary part (one can easily veriy it from Fig. 1 and Fig. 2).
We have used the method of Ganesh and Mishra~\cite{Ganesh:2014lha} to numerically solve 
the one dimensional Schrodinger equation on a logarithamic equally spaced one dimensional lattice. We have obtained the energy 
eigen values and eigen functions of different quarkonium states. 
Then we calculate the energy eigen value at infinity ($U_{\infty}$). 
The binding energy of a given quarkonia state with principal quantum number $n$ and orbital quantum number $l$ is 
calculated by using the following relation~\cite{Satz:2005hx}
\begin{equation}
B. E. (n,l) = g_{nl} - U_{\infty}.
\end{equation}

\begin{figure}
\includegraphics[scale=0.6]{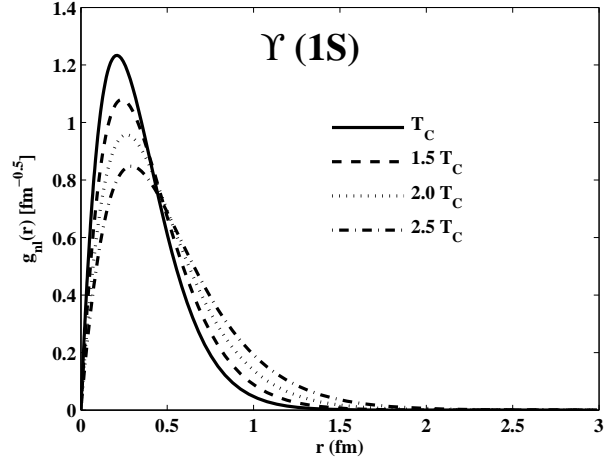}
\caption{Variation of radial wavefunction of $\Upsilon(1S)$ with respect to $r$ 
 at $T=T_{C},~1.5~T_{C},~2.0~T_{C}$ and $2.5~T_{C}$.}
\label{fig7}
\end{figure}
\begin{figure}
\includegraphics[scale=0.6]{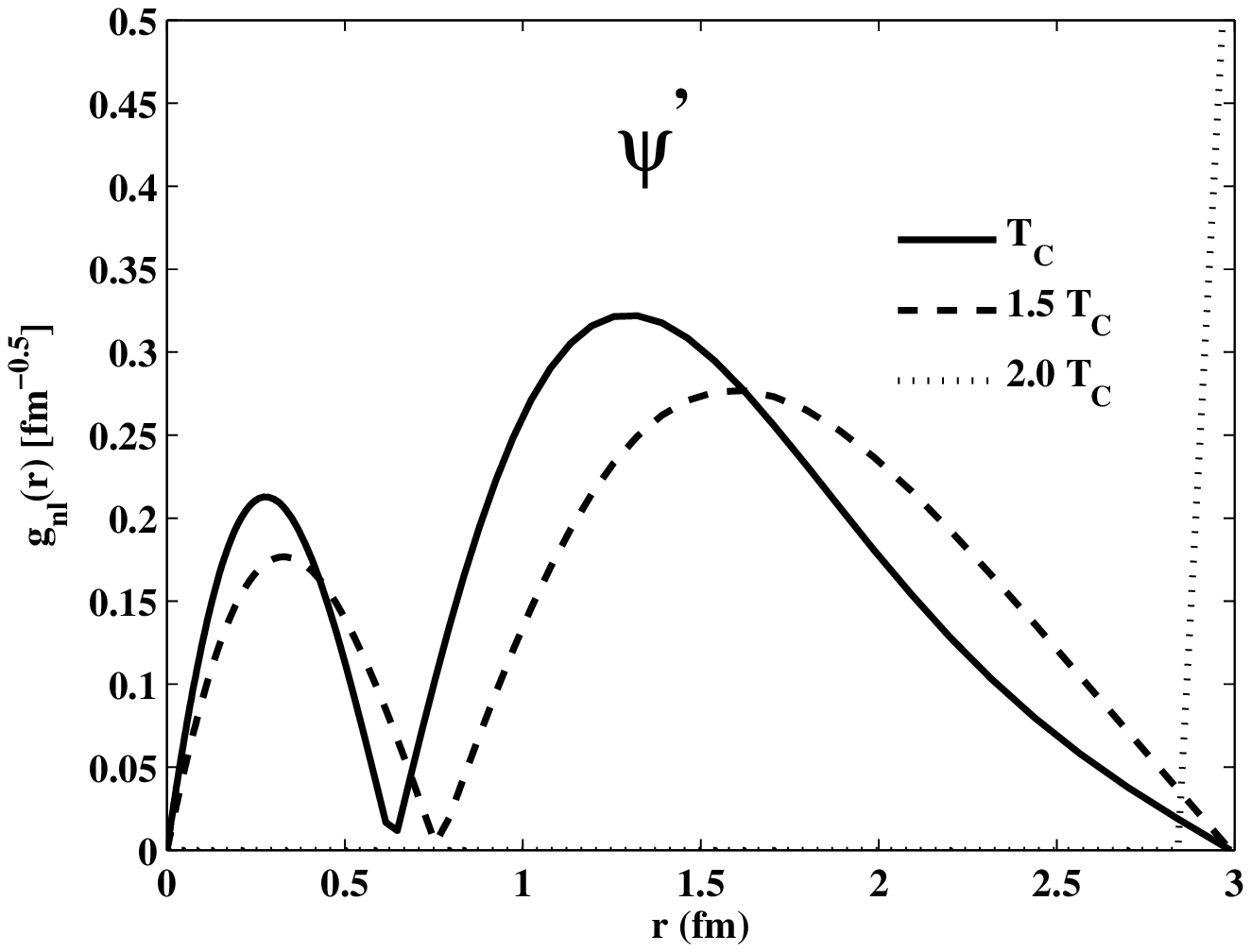}
\caption{Variation of radial wavefunction of $\psi^{'}$ with respect to $r$ at 
 $T=T_{C},~1.5~T_{C},~2.0~T_{C}$ and $2.5~T_{C}$.}
\label{fig8}
\end{figure}
\begin{figure}
\includegraphics[scale=0.6]{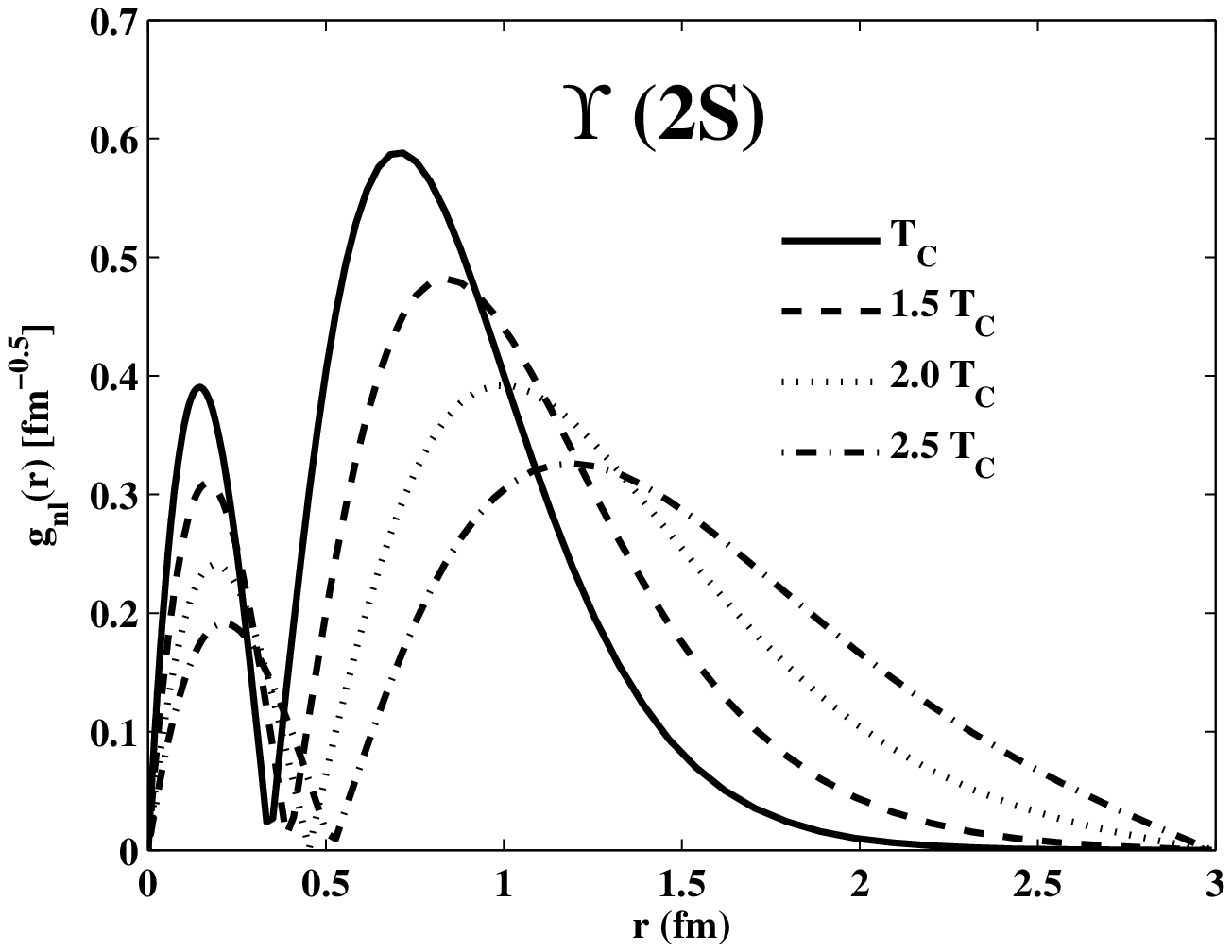}
\caption{Variation of radial wavefunction of $\Upsilon(2S)$ with respect to $r$ 
at $T=T_{C},~1.5~T_{C},~2.0~T_{C}$ and $2.5~T_{C}$.}
\label{fig9}
\end{figure}

\subsection{Decay Width ($\Gamma$)}
\label{C}
\begin{figure}
\includegraphics[scale=0.4]{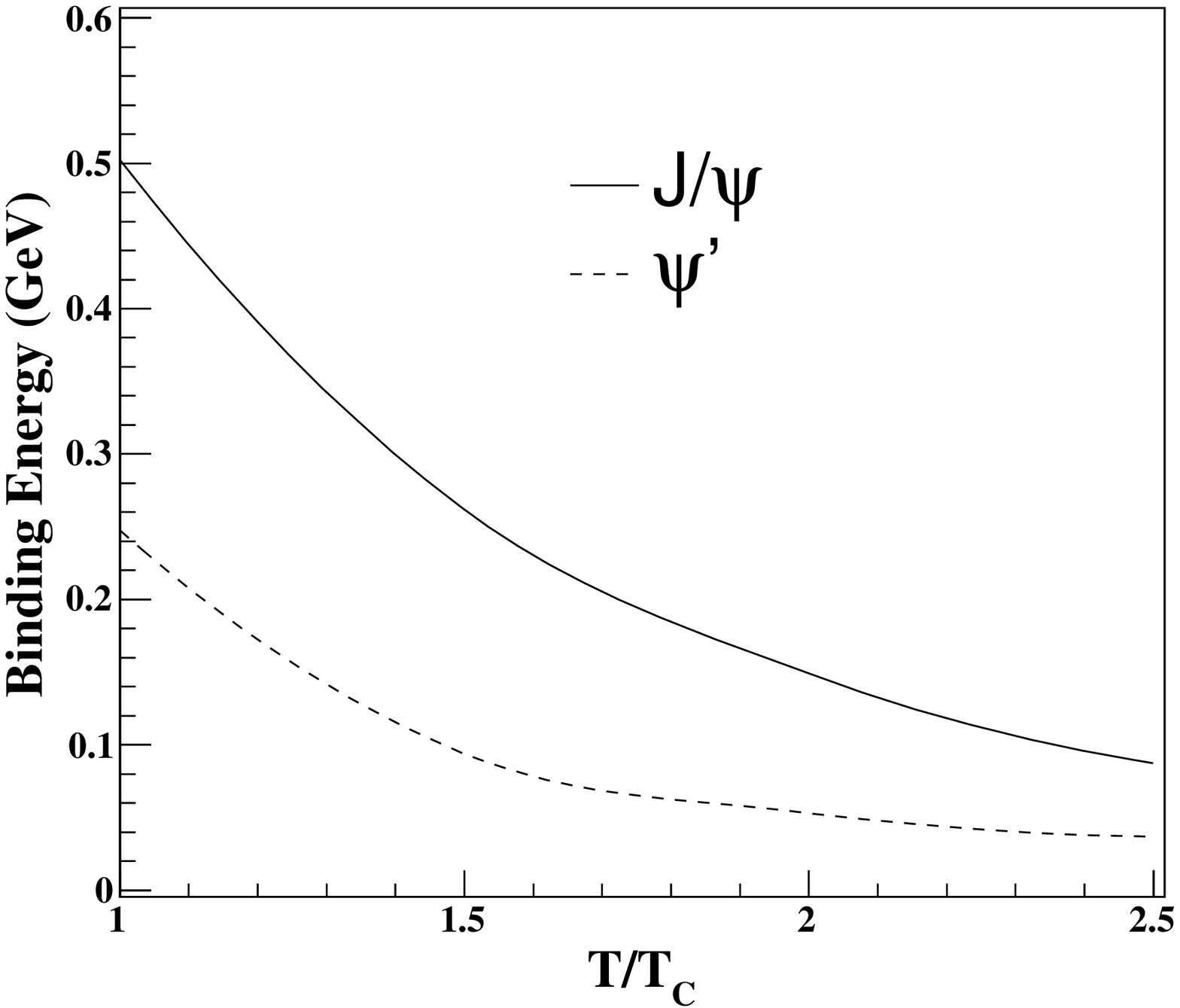}
\caption{Variation of binding energy (B.E.) with respect to temperature in units of $T_{C}$. 
Solid and dashed curves represent the binding energy of $J/\psi$ and $\psi^{'}$, respectively.}
\label{fig10}
\end{figure}

\begin{figure}
\includegraphics[scale=0.4]{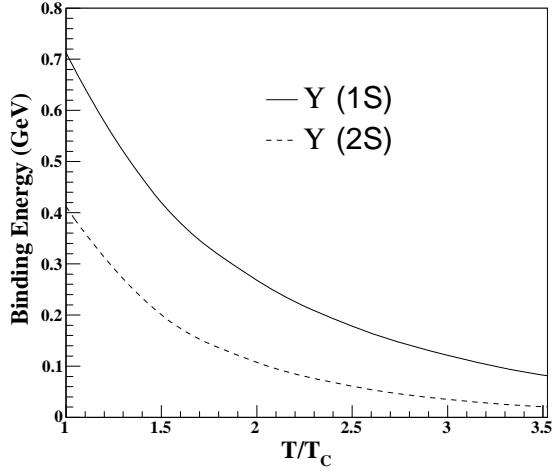}
\caption{Variation of binding energy (B.E.) with respect to temperature in units of $T_{C}$. 
Solid and dashed curves represent the binding energy of $\Upsilon(1S)$ and $\Upsilon(2S)$, respectively.}
\label{fig11}
\end{figure}
We have calculated the 
decay width ($\Gamma$) of $1S$ and $2S$ of quarkonia state numerically. As we know that the thermal width can be calculated from the imaginary part of the potential by using the following expression~\cite{Thakur:2013nia,Thakur:2016cki,Patra:2015qoa}
\begin{equation}
 \Gamma = 4\pi\int g_{nl}^{*}[ImV]g_{nl}~r^{2} dr.
\end{equation}

It is important to mention here that analytically one can calculate the decay width by folding the imaginary part of the potential with 1S and 2S hydrogen atom wavefunction ($g_{_{1S}} = \frac{1}{\sqrt{\pi a_{0}^{3}}} exp\left(-\frac{r}{a_{0}}\right),$ and $g_{_{2S}} = \frac{1}{\sqrt{32\pi a_{0}^{3}}}\left(2-\frac{r}{a_{0}}\right) exp\left(-\frac{r}{2a_{0}}\right)$) which are assumed to represent most of the properties of heavy quarkonia states. However as we will show in our results that the various quarkonia wavefunctions actually depends on temperature very strongly and thus assuming a temperature independent coloumbic wave function to calculate the decay width is not realistic. Thus we have used the wavefunctions as obtained by us, solving Schrodinger equation at different temperatures. 

For the sake of comparison, we are providing here the expression of decay widths obtained by folding the imaginary part of the potential with the coloumbic wavefunctions of different quarkonia state as follows :

\begin{equation}
 \Gamma_{1S} = \frac{4T}{\alpha m_{Q}^{2}}m_{D}^{2}\log\left(\frac{\alpha m_{Q}}{m_{D}}\right)+\frac{4\sigma T}{\alpha^{2}m_{Q}^{2}}\left[1-\frac{3m_{D}^{2}}{\alpha^{2}m_{Q}^{2}}\right]\log\left(\frac{\alpha m_{Q}}{m_{D}}\right),
\end{equation}
and 
\begin{equation}
 \Gamma_{2S} = \frac{56T}{\alpha m_{Q}^{2}}m_{D}^{2}\log\left(\frac{\alpha m_{Q}}{m_{D}}\right)+\frac{8\sigma T}{\alpha^{2}m_{Q}^{2}}\left[7-\frac{192m_{D}^{2}}{\alpha^{2}m_{Q}^{2}}\right]\log\left(\frac{\alpha m_{Q}}{2m_{D}}\right),
\end{equation}
\begin{figure}
\includegraphics[scale=0.4]{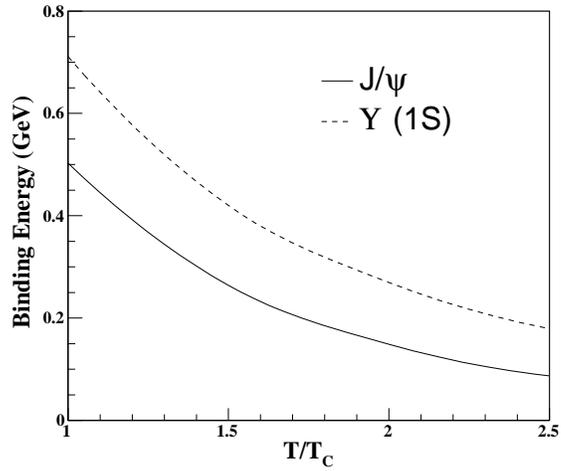}
\caption{Comparison of binding energy of $\Upsilon(1S)$ and $J/\psi$.}
\label{fig12}
\end{figure}
\begin{figure}
\includegraphics[scale=0.4]{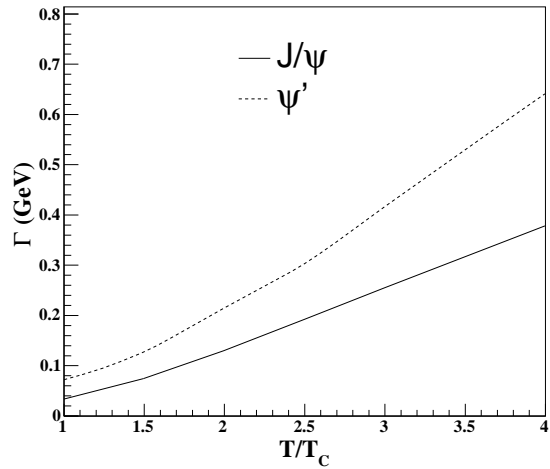}
\caption{Variation of decay width ($\Gamma$) with respect to $T/T_{C}$. Solid and
 dash-dotted curves represent the decay width of $J/\psi$ and $\psi^{'}$, respectively.}
\label{fig13}
\end{figure}

The dissociation temperature for the quarkonium states can be determined by using the conservative quantitative condition
$  \Gamma_{n,~l}(T_{C}) \approx 2\times B. E._{n,~l}(T_{C}) $~\cite{Mocsy:2007jz},
where $T_{C}$ is the dissociation temperature of that particular quarkonia state having principal quantum number 
$n$ and azimuthal quantum number $l$.
\subsection{Survival Probability including Regeneration}
\label{D}
The survival probability or the nuclear modification factor of various quarkonia states with respect to
centrality, rapidity and transverse momentum is key signatures to quantify the properties of medium
created in heavy ion collision experiments. As we have the decay width of various quarkonia
states from our calculation, we can calculate their survival probability by using our recently proposed unified model~\cite{Singh:2015eta}. 
Here we show the variation of 
survival probability with respect to participant number ($N_{part}$) which actually measures the centrality
of collision. To understand the evolution dynamics of the system created in heavy ion collisions, we have used
1+1 dimensional viscous hydrodynamics in which we have included the shear viscous effect. We have done our 
calculation for mid-rapidity region only where Bjorken scaling are applicable. We have derived the cooling 
law for temperature~\cite{Srivastava:2013dxa,Srivastava:2012pd} which depends only on proper time $\tau$ and then extend this cooling law to 
make it $\tau$ and $N_{part}$ dependent using Ref.~\cite{Ganesh:2014lha,Ganesh:2013sqa}. To calculate the survival probability we have used 
the following expression :
\begin{equation}
 S = \frac{N_{HM}^{f}}{N_{HM}^{i}} 
\end{equation}
where $N_{HM}^{i}$ and $N_{HM}^{f}$ is the initial and final multiplicity of heavy meson (quarkonia). The final 
multiplicity of quarkonia can be calculated as~\cite{Singh:2015eta}:
\begin{equation}
 N_{HM}^{f} = \epsilon(\tau_{f})\left[N_{HM}^{i}+N_{Q\bar{Q}}^{2}\int_{\tau_0}^{\tau_f}\Gamma_{f}(V(\tau)\epsilon(\tau))^{-1}d\tau\right].
\end{equation}
Here $\Gamma_{f}$ is the reactivity for the recomination of uncorrelated $Q$ and $\bar{Q}$ quark to form a quarkonia and it 
can be calculated by using decay width~\cite{Singh:2015eta}. 
$N_{Q\bar{Q}}$ is the number density of quark-antiquark pair. $\epsilon(\tau_{f})$ is the dissociation factor which can 
be calculated using the following expression :
\begin{equation}
\epsilon(\tau_f) = exp\left(-\int_{\tau_{0}}^{\tau_{f}}\Gamma~d\tau\right) 
\end{equation}
where $\tau_{0}$ and $\tau_{f}$ are initial and final proper time which actually spans over the QGP lifetime, i.e., 
$\tau_{0}=0.5$ fm and $\tau_{f}=6.0$ fm. We have used $\tau$ and $N_{part}$ dependent cooling law for temperature
as follows~\cite{Srivastava:2013dxa,Ganesh:2013sqa}:
\begin{equation}
T(\tau) = T_{c}\left(\frac{N_{part}(bin)}{N_{part}(bin_{0})}\right)^{1/3}\left(\frac{\tau_{QGP}}{\tau}\right)^{1/3},
\end{equation}
where $N_{part}(bin_{0})$ is the number of participant corresponding to the most central bin as used in our 
calculation and $N_{part}(bin)$ is the number of participant corresponding to the bin at which we want to
calculate the temperature. $\tau_{QGP}$ is the lifetime of QGP.\\

The cooling law for volume is derived using the condition of isentropic evolution of the medium and can be expressed 
as follows~\cite{Singh:2015eta}:
\begin{equation}
 V(\tau,~b) = V(\tau_0,~b)\left(\frac{\tau_{0}}{\tau}\right)^{\left(\frac{1}{R} -1\right)}
\end{equation}
where $V(\tau_0,~b)= \pi\;(r_{t} - b/2)^{2}\tau_{0}$ is volume at the initial time $\tau_0$ and an impact of $b$ fm.\\
$N_{c\bar{c}}$ and $N_{b\bar{b}}$ are calculated in our model using the help of Glauber model. The extrapolation to the nucleus-nucleus 
collisions is done via standard overlap integral scaling as follows :
\begin{equation}
 N_{c\bar{c}} (b) = \sigma_{c\bar{c}}^{NN}\; T_{AA}
\end{equation}
where $\sigma_{c\bar{c}}^{NN}$ is the cross section for $c\bar{c}$ pair
production in p$+$p collision. The $\sigma_{c\bar{c}}^{NN}$ has been calculated
using pQCD approach for GRV HO hadronic structure function~\cite{Thews:2005fs}, we
have obtained $\sigma_{c\bar{c}}^{NN} = 3.546$ mb and $\sigma_{b\bar{b}}^{NN} = 0.1105$ mb for LHC at $\sqrt{s} = 2.76$
TeV. Further we have obtained $\sigma_{c\bar{c}}^{NN} = 0.346$ mb $\sigma_{b\bar{b}}^{NN} = 0.01035$ mb for RHIC at  $\sqrt{s} = 200$ GeV.
Here, $T_{AA}(b)$ is nuclear overlap function, its impact parameter ($b$)
dependent values have been taken from Ref.~\cite{cern}. Here it is important to mention that we have not 
incorporated any type of cold nuclear matter (CNM) effect in the present calculations.
\section{Results and Discussions}
\label{secIII}
\begin{figure}
\includegraphics[scale=0.4]{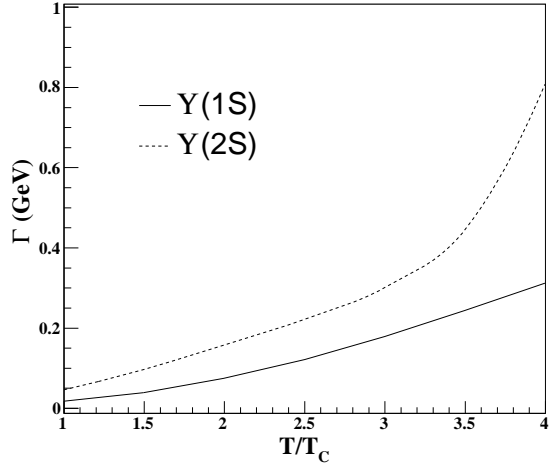}
\caption{Variation of decay width ($\Gamma$) with respect to $T/T_{C}$. Solid and
 dashed curves represent the decay width of $\Upsilon(1S)$ and $\Upsilon(2S)$, respectively.}
\label{fig14}
\end{figure}
\begin{figure}
\includegraphics[scale=0.4]{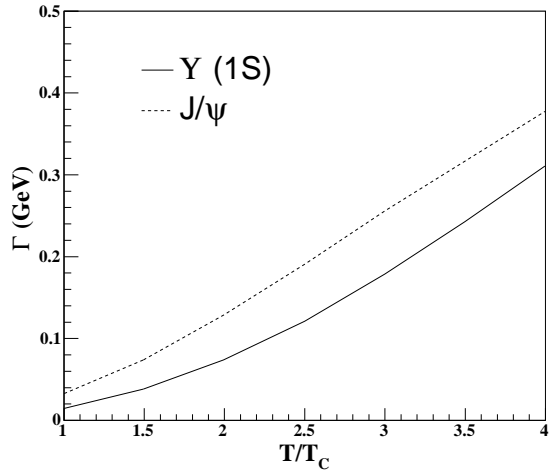}
\caption{Comparison of decay width of $J/\psi$ and $\Upsilon(1S)$.}
\label{fig15}
\end{figure}
\begin{figure}
\includegraphics[scale=0.4]{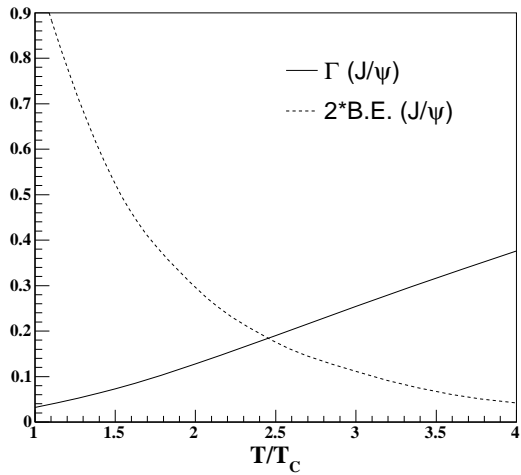}
\caption{Variation of two times of binding energy of $J/\psi$ and decay width with respect to $T/T_{C}$.}
\label{fig16}
\end{figure}
The main ingredient of this paper is the heavy quark potential in QCD plasma~\cite{Agotiya:2008ie,Thakur:2013nia}. 
We first show few characteristics of this potential. In Fig. \ref{ReV}, we demonstrate 
the variation of the real part of the heavy-quark potential with respect to the separation
distance ($r$) between 
the $ Q\bar{Q} $ pair. We have plotted the real part of potential at three different values of temperature, i.e., $T=160,~320$ and $480$ MeV by solid, dashed and dash-dotted curve, respectively.
The real potential starts 
from negative value and increases very sharply to zero as we increase the 
distance from zero to $0.5$ fermi. Further the real part of potential increases
from zero to $0.4$ GeV as we increase the distance at $T=160$ MeV. We choose 
this specific value of temperature since we take $160$ MeV as the critical 
crossover temperature ($T_{C}$) in our calculation. We also show the variation
in the saturation value of real potential with increase in temperature. As we 
increase the temperature the potential saturates at lower values, i.e., $0.4$ GeV
at $T=160$ MeV to $0.05$ GeV at $T=480$ MeV. In Fig. \ref{ImV}, we have plotted the imaginary part of heavy-quark potential with respect to $r$. The imaginary
potential starts from zero value at $r=0$ fm and then decreases and became negative
with increase in $r$. As we increase the temperature the magnitude of imaginary 
potential also increases in the negative direction. At higher temperatures, we observed a flucuating behaviour for $r > 0.4$ fm. This is due to the $sine$ term in imaginary potential.
 
Fig.~\ref{fig3} represents the variation of radial part of eigen wavefunction 
for a given state $n$ and $l$, i.e., $g_{nl}(r)$ of $J/\Psi$ and $\Psi^{'}$ 
and $\chi_{C}$ states with respect to the $r$ at the critical 
temperature, $T_{C}$. Similarly, Fig.~\ref{fig4} demonstrates the radial part of wavefunction of different
bottomonia states and their variation with respect to $r$ at $T_{C}$.

\begin{figure}
\includegraphics[scale=0.4]{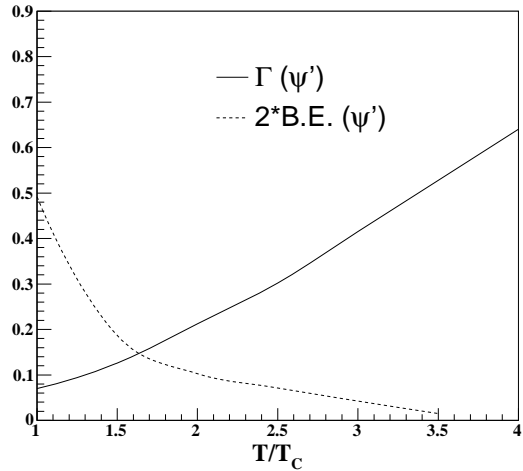}
\caption{Variation of two times of binding energy of $\psi^{'}$ and decay width with respect to $T/T_{C}$.}
\label{fig17}
\end{figure}
\begin{figure}
\includegraphics[scale=0.4]{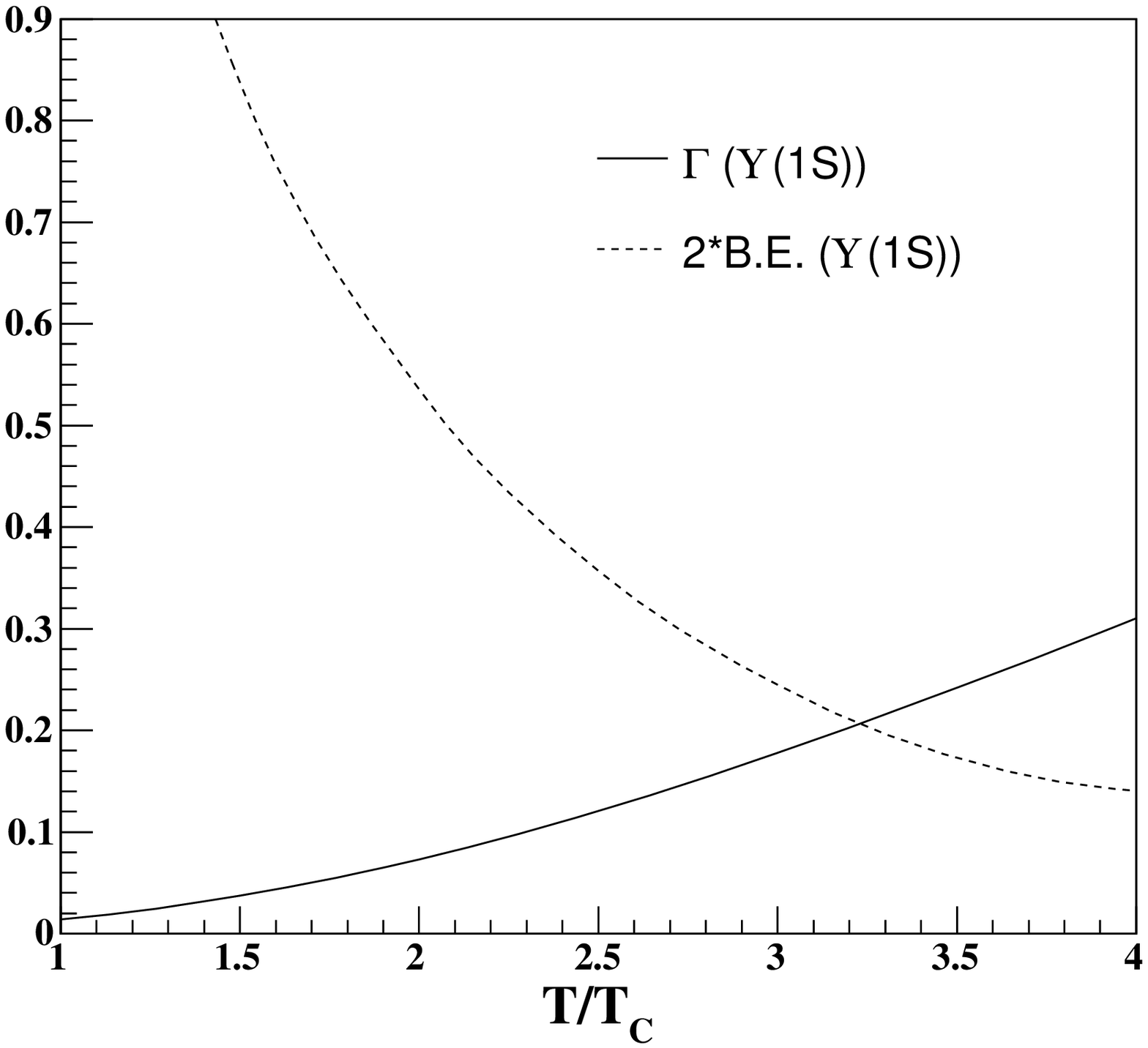}
\caption{Variation of two times of binding energy of $\Upsilon(1S)$ and decay width with respect to $T/T_{C}$.}
\label{fig18}
\end{figure}
\begin{figure}
\includegraphics[scale=0.4]{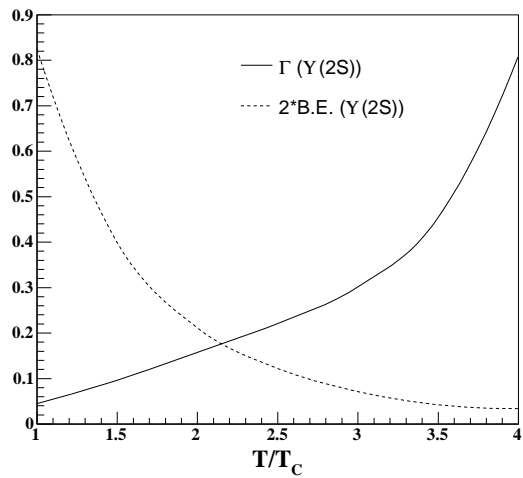}
\caption{Variation of two times of binding energy of $\Upsilon(2S)$ and decay width with respect to $T/T_{C}$.}
\label{fig19}
\end{figure}
Fig.~\ref{fig5} presents a comparison between eigen wave function of charmonium ($J/\Psi$) 
and bottomonium ($\Upsilon(1S)$). Here one can clearly see the difference in 
peak height and peak-width of $J/\Psi$ and $\Upsilon(1S)$ and understand the 
strong binding of $b-\bar{b}$ quark in $\Upsilon(1S)$ in comparison to the 
binding of $c-\bar{c}$ quark in $J/\Psi$. 

Fig.~\ref{fig6} shows the change in the radial part of wavefunction of charmonium as we increase the temperature
from $T_{C}$ to $2.5~T_{C}$ in the step of $0.5~T_{C}$. From here it is clear that as we 
increase the temperature the peak-height of eigen function decreases and the peak-width
increases which causes the binding between the heavy quark and anti-quark in the 
the bound state of charmonium to decrease. Similarly we have shown the change in 
eigen function of $\Upsilon(1S)$ with respect to temperature in Fig.~\ref{fig7}.

\begin{figure}
\includegraphics[scale=0.4]{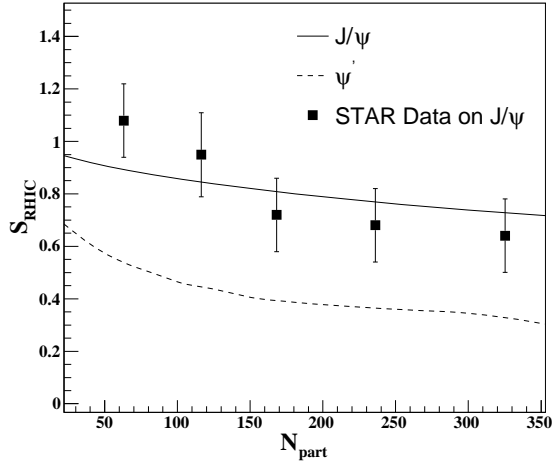}
\caption{Variation of survival probability (S) of $J/\psi$ and $\psi^{'}$ with respect to $N_{part}$ at 
center of mass energy $\sqrt{s_{NN}}=200$ GeV. Experimental Data is taken from Ref.~\cite{Adamczyk:2012ey}.}
\label{fig20}
\end{figure}
\begin{figure}
\includegraphics[scale=0.4]{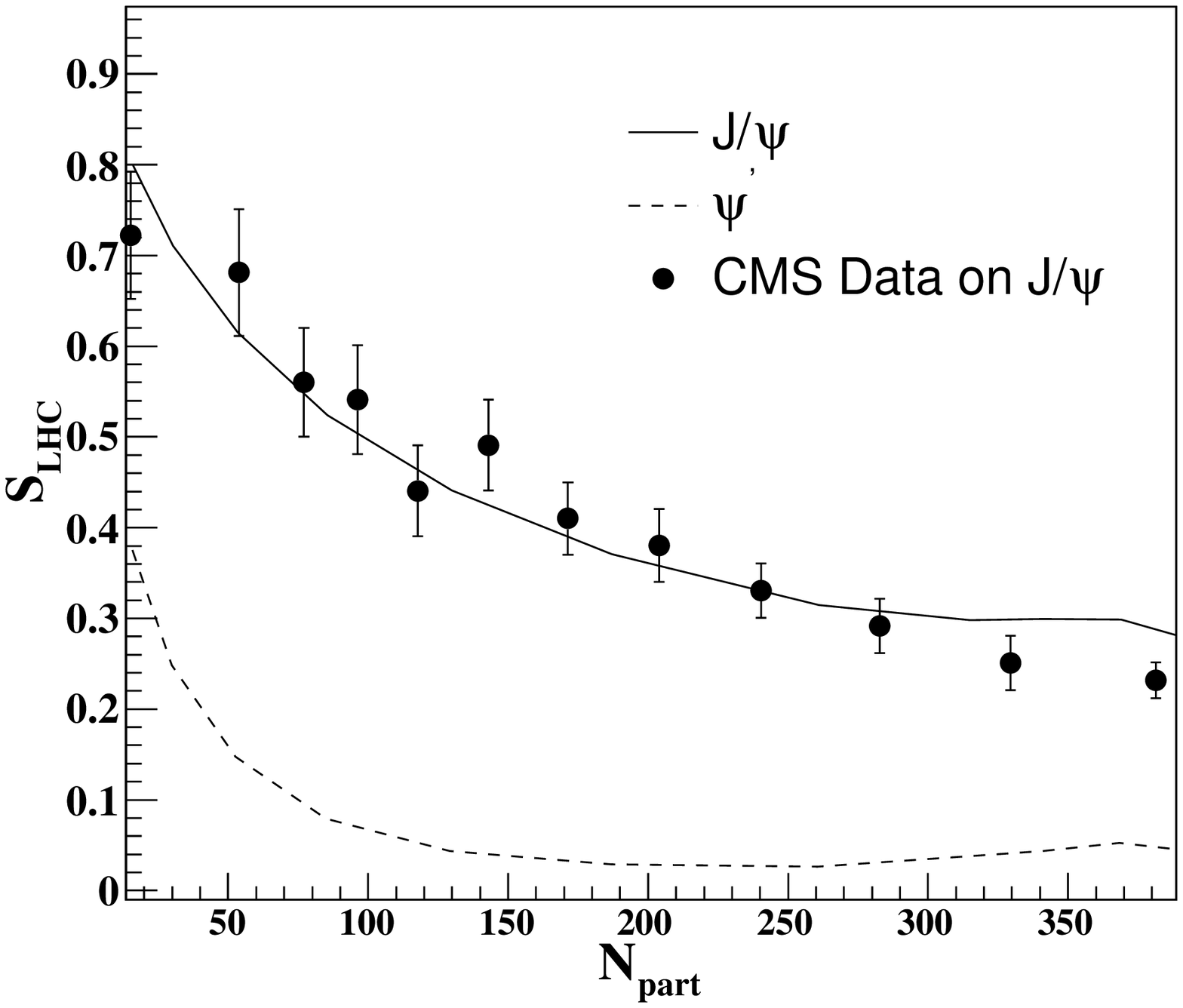}
\caption{Variation of survival probability (S) of $J/\psi$ and $\psi^{'}$ with respect to $N_{part}$ at 
center of mass energy $\sqrt{s_{NN}}=2.76$ TeV. Experimental Data is taken from Ref.~\cite{Murray:2012fya}.}
\label{fig21}
\end{figure}
\begin{figure}
\includegraphics[scale=0.4]{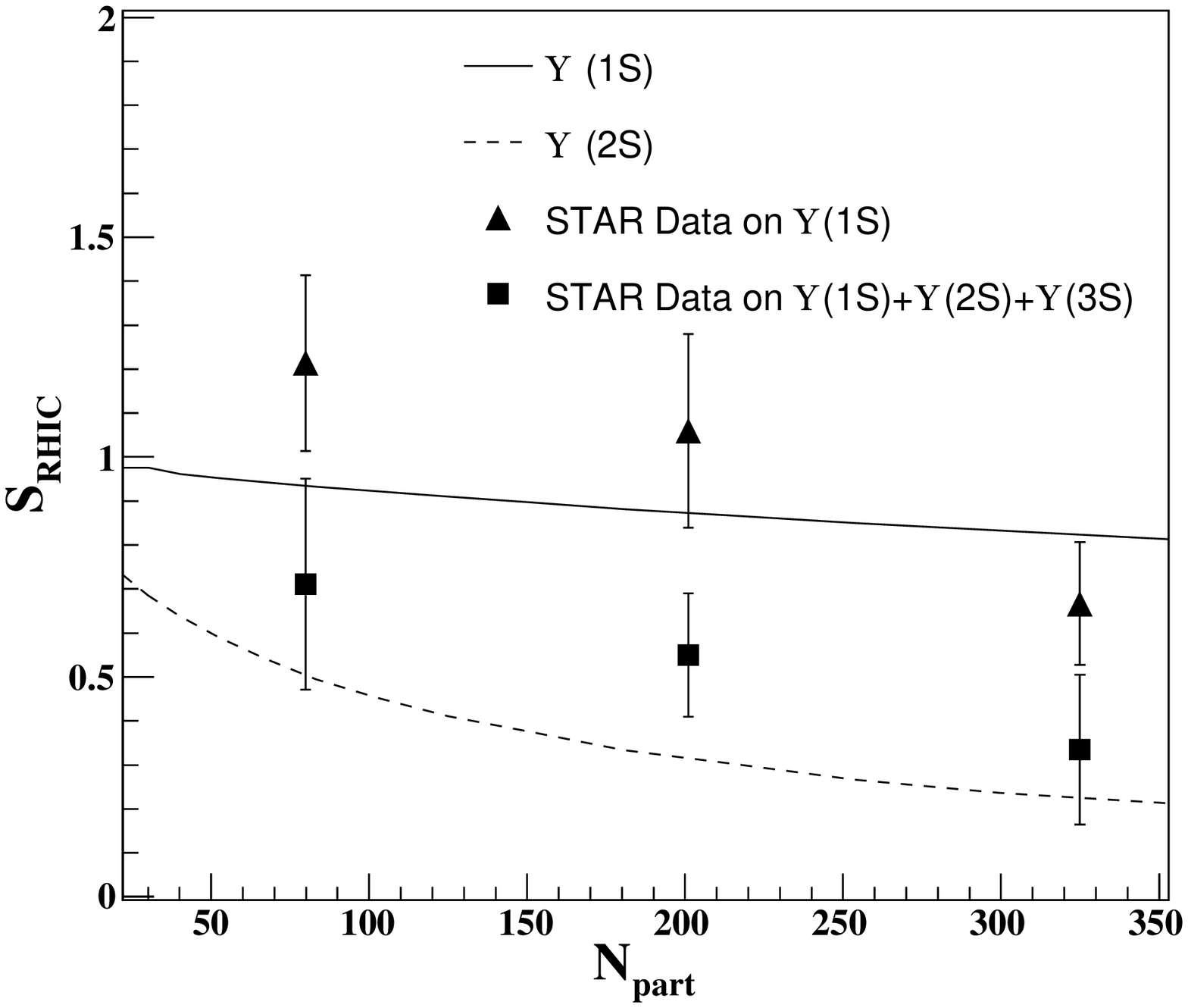}
\caption{Variation of survival probability (S) of $\Upsilon(1S)$ and $\Upsilon(2S)$ with respect to $N_{part}$ at 
center of mass energy $\sqrt{s_{NN}}=200$ GeV. Experimental Data are taken from Ref.~\cite{Adamczyk:2014ey}.}
\label{fig22}
\end{figure}

\begin{figure}
\includegraphics[scale=0.4]{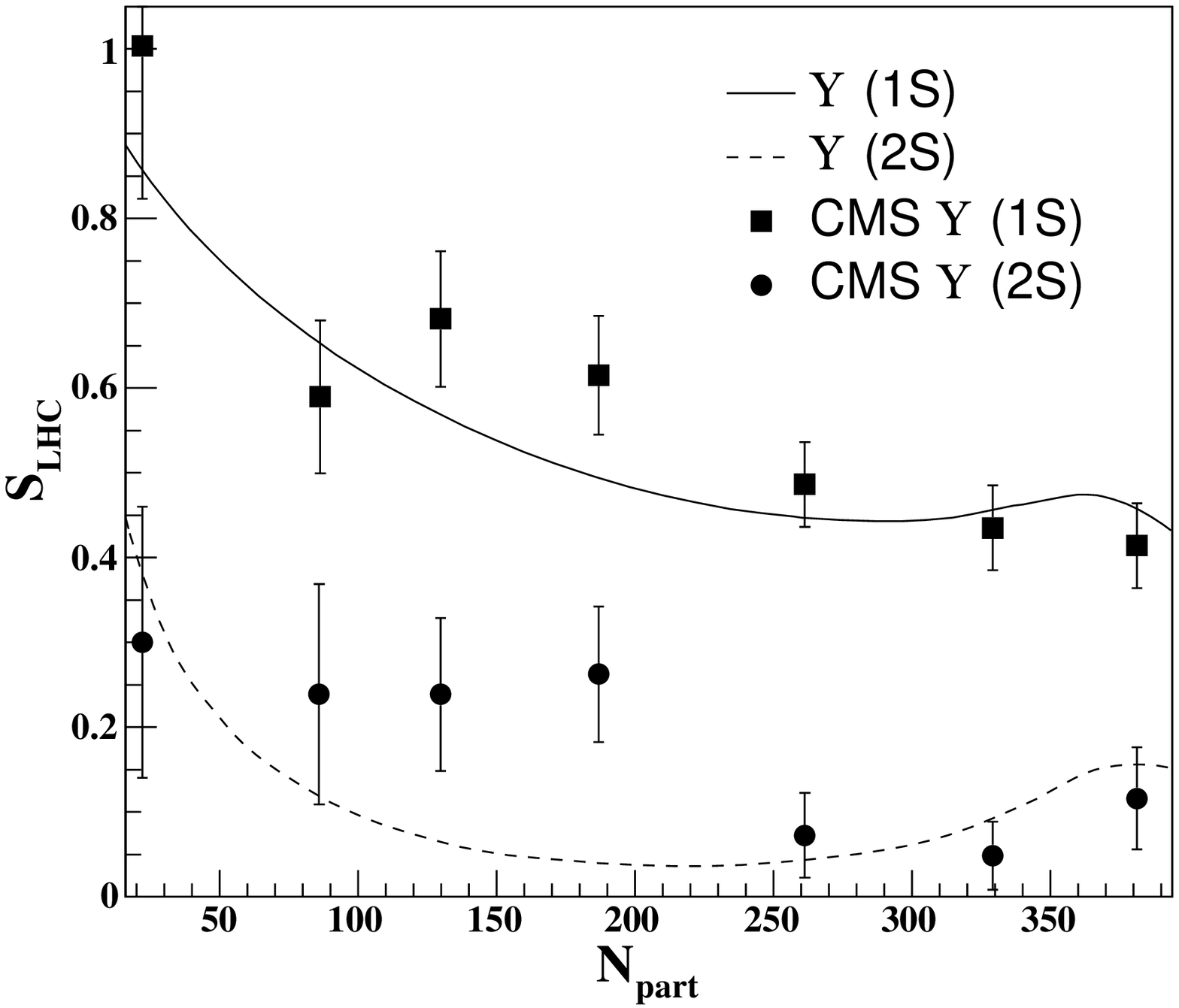}
\caption{Variation of survival probability (S) of $\Upsilon(1S)$ and $\Upsilon(2S)$ with respect to $N_{part}$ at 
center of mass energy $\sqrt{s_{NN}}=2.76$ TeV. Experimental Data are taken from Ref.~\cite{Murray:2012fya}.}
\label{fig23}
\end{figure}

\begin{figure}
\includegraphics[scale=0.4]{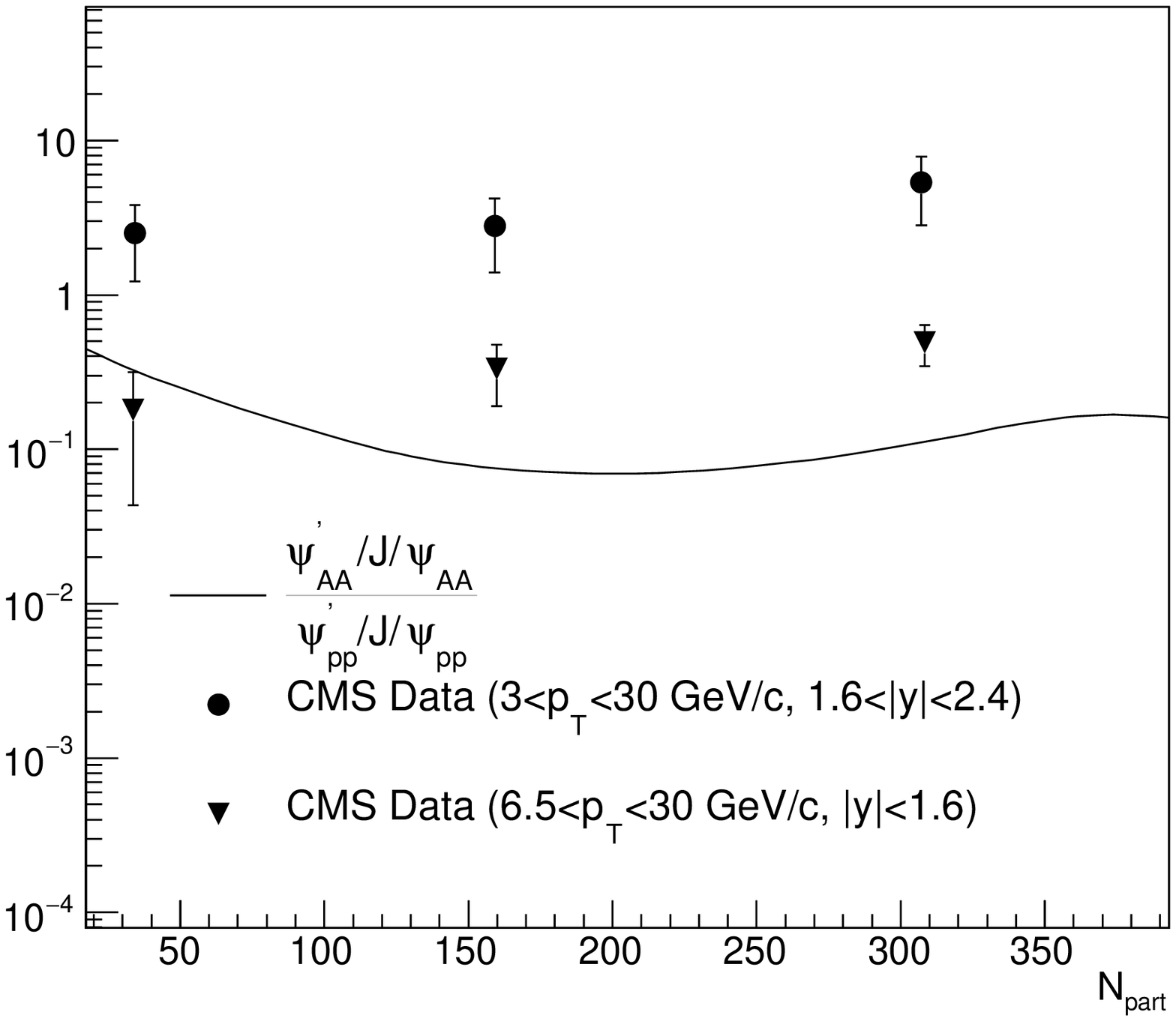}
\caption{Variation of survival probability (S) of $[\psi^{'}/J/\psi]_{PbPb}/[\psi^{'}/J/\psi]_{pp}$ with respect to $N_{part}$ at 
center of mass energy $\sqrt{s_{NN}}=2.76$ TeV. Experimental Data are taken from Ref.~\cite{Murray:2012fya}.}
\label{fig24}
\end{figure}
In Fig.~\ref{fig8} and \ref{fig9}, we demonstrate the change in shape of eigen functions of  $\Psi^{'}$ 
and $\Upsilon(2S)$ state with respect to temperature going from $T_{C}$ to $2.5~T_{C}$, respectively.
Fig.~\ref{fig10} represents the variation of binding energy of $J/\Psi$ and $\Psi^{'}$ with respect to
temperature. Here we present the temperature in the unit of $T_{C}$. It is clear from
the plot that initially when the temperature is near $T_{C}$, the binding energy of $J/\Psi$ is 
large and thus charmonium can still survive after critical temperature. As we start to increase 
the temperature from $T_{C}$ to higher values, the binding energy of charmonium starts decreasing 
and acquires a low value which is near to zero. However, our calculation shows that even 
at $2.5~T_{C}$, there is a finite value of binding energy for $J/\Psi$. For $\Psi^{'}$, the 
binding energy starts from a lower value in comparison to $J/\Psi$ which is quite obvious and 
it decreases with increase in temperature and acquire almost zero value at $2.0~T_{C}$. 

Bottomonia wavefunction is more columbic at $T_{C}$ in comparison to charmonia states due to the large mass of $\Upsilon(1S)$ and $\Upsilon(2S)$ in comparison to $J/\Psi$ and $\Psi^{'}$. 
Therefore the binding energy of various bottomonia states start from a higher value in comparison 
to corresponding charmonia states. Fig.~\ref{fig11} represents the variation of binding energy of $\Upsilon(1S)$
 and $\Upsilon(2S)$ with respect to $T/T_{C}$.  Further, Fig.~\ref{fig12} shows the comparison of 
 binding energy of charmonia $1S$ state with bottomonia $1S$ state.

We now present the decay width coming only due to the imaginary 
part of the heavy quark potential. In Fig.~\ref{fig13}, we demonstrate the variation of decay width of 
1S and 2S charmonia states with respect to $T/T_{C}$. As we have earlier shown in Fig.~\ref{fig10}
that the binding energy of $J/\Psi$ is large at $T_{C}$ and thus the decay width should
be small at $T_{C}$. As the binding energy of $\Psi^{'}$ 
is less in comparison to $J/\Psi$ over the entire temperature range therefore $\Psi^{'}$ has a larger decay width than $J/\Psi$ as shown in this figure at each temperature. Further, the decay width increases with the increase in temperature. Furthermore, the difference between the decay width of  $J/\Psi$ and $\Psi^{'}$ increases with the temperature.
\begin{center}
\begin{table}
\caption{Dissociation Temperature ($T_{d}$) for various quarkonium states.}
  \begin{tabular}{c c c c}
    \hline
    Charmonium States \vline & Diss. Temp. ($T_{d}$)\vline &Bottomonium States \vline &Diss. Temp. ($T_{d}$)\\
    \hline
    $J/\psi$  & $2.4~T_{C}$ & $\Upsilon(1S)$ & $3.2~T_{C}$\\
    \hline
    $\psi^{'}$ & $1.6~T_{C}$ & $\Upsilon(2S)$ & $2.2~T_{C}$\\
   \hline
  \end{tabular}
  \end{table}
\end{center}
In Fig.~\ref{fig14}, we have plotted the variation of decay width of $\Upsilon(1S)$ and $\Upsilon(2S)$ with
respect to $T/T_{C}$. The trend is quite similar with the charmonia states but the decay width of $\Upsilon(2S)$ increases rapidly after $T=3.5~T_{C}$. Fig.~\ref{fig15} presents a comparison
between the decay width of $J/\Psi$ and $\Upsilon(1S)$. This figure clearly shows that decay
width of $J/\Psi$ and $\Upsilon(1S)$ starts from almost similar value at $T_{C}$. However, the width of
$J/\Psi$ increases more rapidly in comparison to $\Upsilon(1S)$. This means 
that the $J/\Psi$ dissociates at lower temperatures in comparison to bottomonium which is a tightly bound
state and thus survive to higher temperatures.\\

After that we have obtained the dissociation temperature for the different quarkonium states. 
 Different 
dissociation criteria have been discussed in the literature. The first criteria is that a quarkonium state
should dissociate at the temperature $T$ where $B.E. = T$. Here $B.E.$ is the binding 
energy of that particular quarkonia state. This criteria can provide an upper bound on the dissociation
temperature. Here we use a more strict dissociation criteria which suggest that any quarkonium state 
should dissociate at that temperature where the decay width of the quarkonium state becomes equal to two times 
of its binding energy, i.e., $\Gamma=2~B.E.$~\cite{Mocsy:2007jz,Thakur:2013nia}.

In Fig.~\ref{fig16} we have plotted the decay width and two times of binding energy of $J/\Psi$ with respect 
to $T/T_{C}$. The two curves intersect each other at $2.4~T_{C}$. Thus the dissociation temperature of 
$J/\Psi$ comes out as $2.4~T_{C}$ in our calculation. We have plotted the width and the two times of binding energy of $\Psi^{'}$ with respect to $T/T_{C}$ in Fig.~\ref{fig17}. The dissociation temperature of $\Psi^{'}$ as obtained from the graph is $1.6~T_{C}$.  



Figs.~\ref{fig18} and \ref{fig19} are the similar graphs but for 
$\Upsilon(1S)$ and $\Upsilon(2S)$ states, respectively. The dissociation temperature as obtained from the respective Figs.~\ref{fig18} and \ref{fig19}
are $3.2~T_{C}$ and $2.2~T_{C}$. 

In Fig.~\ref{fig20}, we present the variation of survival probability of $J/\Psi$ and $\Psi^{'} $with respect to $N_{part}$ 
at highest RHIC energy, i.e., $\sqrt{s_{NN}}=200$ GeV. As it is clear from the binding energy and decay width
curve of $J/\psi$ and $\psi^{'}$, the dissociation probability of $\psi^{'}$ is large in comparison to $J/\psi$. Thus 
the survival probability of $J/\psi$ is larger than the $\psi^{'}$ at each centrality. In other words, it means a less suppression of $J/\psi$ in comparison to $\psi^{'}$. We have compared our $J/\psi$ results with the corresponding results obtained by STAR experiment~\cite{Adamczyk:2012ey}. Our model results underestimate the data in most peripheral collisions. However, it suitably describes the data for central and semi-peripheral collisions. Fig.~\ref{fig21} shows the variation of
survival probability of $J/\psi$ and $\psi^{'}$ with respect to centrality at LHC energy, i.e., $\sqrt{s_{NN}}=2.76$ TeV.
As the energy increases, the corresponding temperature and energy density in each centrality class also increases and
thus the survival probability of $J/\psi$ and $\psi^{'}$ decreases in comparison to RHIC energy results.  We have compared our model result with the experimental data obtained by CMS Collaboration~\cite{Murray:2012fya}. Model results for $J/\psi$ satisfy the data over the entire centrality region except in extreme central collisions. From this plot, one can observe that as we move towards the central collisions from the semiperipheral collisions, there is a small rise in the survival probability. This rise is quite clearly visible in the case of $\psi^{'}$. The rise of survival probability in central collsions at LHC energy is due to the rise of regeneration effect through recombination of charm-anticharm pairs in the later stage of the medium evolution. We found in our calculation that the regeneration effect is negligibly small at RHIC energy.
\begin{figure}
\includegraphics[scale=0.4]{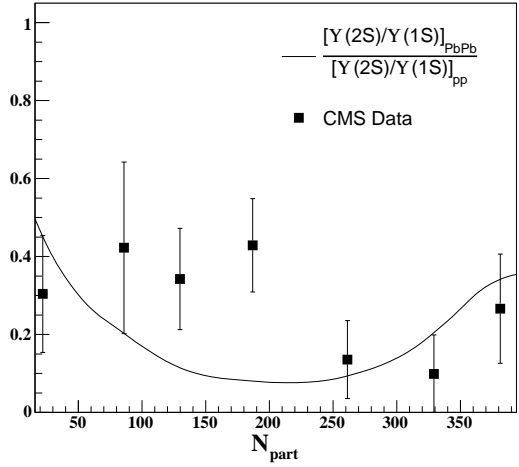}
\caption{Variation of survival probability (S) of $[\Upsilon(2S)/\Upsilon(1S)]_{PbPb}/[\Upsilon(2S)/\Upsilon(1S)]_{pp}$ with respect to $N_{part}$ at 
center of mass energy $\sqrt{s_{NN}}=2.76$ TeV. Experimental Data is taken from Ref.~\cite{Murray:2012fya}.}
\label{fig25}
\end{figure}
Figures \ref{fig22} and \ref{fig23}
 presents the survival probability of $\Upsilon(1S)$ and $\Upsilon(2S)$ at RHIC and LHC energies, respectively. Due to
their large mass in comparison to charmonia, the decay width is small and thus the survival probability of
$\Upsilon(1S)$ and $\Upsilon(2S)$ is large at each $N_{part}$ in comparison to the survival probability of $J/\psi$
and $\psi^{'}$, respectively. We have compared our model results with the experimental data wherevere they are available. In Fig.~\ref{fig22}, we have shown the STAR data~\cite{Adamczyk:2014ey} for $\Upsilon(1S)$ and combined suppression of $\Upsilon(1S)+\Upsilon(2S)+\Upsilon(3S)$. Further in Fig.~\ref{fig23}, we have plotted the CMS data~\cite{Murray:2012fya} for $\Upsilon(1S)$ and $\Upsilon(2S)$. We observed that our model at RHIC energy is able to reproduce the experimental data of $\Upsilon(1S)$. Further at LHC energy, model results regarding $\Upsilon(1S)$ and $\Upsilon(2S)$ satisfy the experimental data satisfactorily. However our results underestimate the data in semi-peripheral collisions. In central collisions, one can again observe the clear effect of regeneration on the survival probability of $\Upsilon (1S)$ and $\Upsilon(2S)$ which is negligibly small at RHIC energy.

Finally we have calculated the double ratio between charmonia states ($\psi^{'}$ and $J/\psi$)  and bottomonia states ($\Upsilon(2S)$ and $\Upsilon(1S)$) at $2.76$ TeV with respect to $N_{part}$ in Figs.~\ref{fig24} and \ref{fig25}, respectively. In Fig.~\ref{fig24}, we have compared model result for the double ratio of charmonium states with the recent experimental data from CMS Collaboration at mid-rapidity as well as at forward rapidity. Since we have done our calculation for survival probability at mid-rapidity thus one can observe that our model with modified heavy quark potential satisfies the data well in most peripheral collisions while it underestimates the data at semicentral and central collisions. Similarly, in Fig.~\ref{fig25}, we have compared our model results for the double ratio of bottomonium states with the experimental data. We observed that the model results satisfy the data satisfactorily well. Only in the semi-peripheral region, the model underestimates.

In summary, we have solved the 1+1 dimension Schrodinger equation using a modified heavy quark potential and obtained the eigen function and eigen values of the different charmonium and bottomonium states. We have also calculated the binding energy by solving the 1+1 dimensional Schrodinger equation for an infinite heavy quark potential. Further we have obtained the decay width of different quarkonium states using temperature dependent wavefunction obtained by us and demonstrate their variation with respect to temperature. Furthermore we have obtained the dissociation temperature of diffferent quarkonium states by using the dissociation criteria, i.e., decay width = $2\times$binding energy. After that we feed these decay widths in our recently proposed unified model to calculate the survival probability of various quarkonium states. We have also obtained the nuclear modification factor for double ratios and observed that our model suitably describes the experimental data regarding nuclear modification factor (or survival probability).

\noindent
\section{Acknowledgments}
PKS acknowledges IIT Ropar, India for providing an institute postdoctoral research grant.

\newpage
\end{document}